\documentclass[fleqn,10pt]{wlscirep}
\usepackage[utf8]{inputenc}
\usepackage[T1]{fontenc}
\usepackage{subcaption}
\usepackage{bm}
\usepackage{arydshln}
\usepackage{enumitem}
\def\CR{\text{COVID-Rate}}
\def\CRs{\text{COVID-CT-Rate}}

\title{$\CR$: An Automated Framework for Segmentation of COVID-19 Lesions from Chest CT Scans}

\author[1]{Nastaran Enshaei}
\author[2,*]{Anastasia Oikonomou}
\author[3]{Moezedin Javad Rafiee}
\author[1]{Parnian Afshar}
\author[4]{Shahin Heidarian}
\author[1]{Arash Mohammadi}
\author[5]{Konstantinos N. Plataniotis}
\author[1]{Farnoosh Naderkhani}
\affil[1]{Concordia Institute for Information Systems Engineering, Concordia University, Montreal, QC, Canada.}
\affil[2]{Department of Medical Imaging, Sunnybrook Health Sciences Centre, University of Toronto, Toronto, ON, Canada.}
\affil[3]{Department of Medicine and Diagnostic Radiology, McGill University, Montreal, QC, Canada.}
\affil[4]{Department of Electrical and Computer Engineering, Concordia University, Montreal, QC, Canada}
\affil[5]{Department of Electrical and Computer Engineering, University of Toronto, Toronto, ON, Canada.}

\affil[*]{anastasia.oikonomou@sunnybrook.ca}

\begin{abstract}
Novel Coronavirus disease (COVID-19) is a highly contagious respiratory infection that has had devastating effects on the world. Recently, new COVID-19 variants are emerging making the situation more challenging and threatening. Evaluation and quantification of COVID-19 lung abnormalities based on chest Computed Tomography (CT) scans can help determining the disease stage, efficiently allocating limited healthcare resources, and making informed treatment decisions. During pandemic era, however, visual assessment and quantification of COVID-19 lung lesions by expert radiologists become expensive and prone to error, which raises an urgent quest to develop practical autonomous solutions. In this context,  first, the paper introduces an open-access COVID-19 CT segmentation dataset containing $433$ CT images from $82$ patients that have been annotated by an expert radiologist. Second, a Deep Neural Network (DNN)-based framework is proposed, referred to as the $\CR$, that autonomously segments lung abnormalities associated with COVID-19 from chest CT scans.  Performance of the proposed $\CR$ framework is evaluated through several experiments based on the introduced and external datasets. The results show a dice score of $0.802$ and specificity and sensitivity of $0.997$ and $0.832$, respectively. Furthermore, the results indicate that the $\CR$ model can efficiently segment COVID-19 lesions in both 2D CT images and whole lung volumes. Results on the external dataset illustrate generalization capabilities of the $\CR$ model to CT images obtained from a different scanner.
\end{abstract}
\begin{document}

\flushbottom
\maketitle
\thispagestyle{empty}

\section*{Introduction}
Coronavirus disease 2019 (COVID-19) has been the world's most threatening challenge of the twenty first century. According to the Coronavirus Resource Center of John Hopkins University (JHU)~\cite{JHU2021}, over $179$ million confirmed cases of COVID-19, including over $3.8$ million deaths, have been reported in $192$ countries/regions by 24 June 2021. Numerous studies indicate that COVID-19 vaccination can effectively reduce disease transmission, hospital admissions, and deaths. However, due to new emerging COVID-19 variants~\cite{NavecaF2021, WangR2021, JassatW2021} the number of COVID-19 daily cases is still increasing in many areas. It is, therefore, crucial for healthcare professionals and authorities across the globe to find practical solutions to manage the COVID-19 pandemic and learn from this experience to be well prepared for potential future ones.

\begin{figure}[t!]
\centering
\includegraphics[scale=0.8]{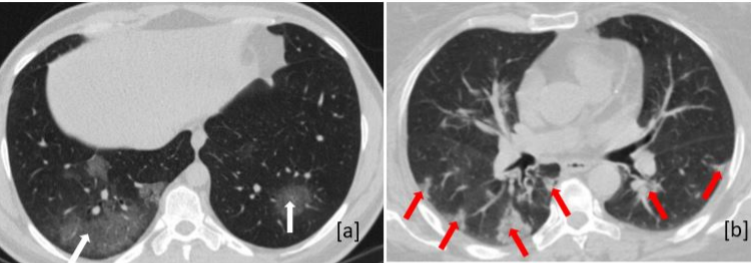}
\caption{\small The most commonly observed CT patterns in COVID-19 pneumonia. (a) GGO pattern demonstrated with white arrows, (b) Consolidation pattern demonstrated with red arrows.~\cite{mohammadi2020diagnosis}}
\label{fig:lunginf}
\end{figure}

Reverse Transcription-Polymerase Chain Reaction (RT-PCR) is the gold-standard test for the diagnosis of COVID-19. However, the RT-PCR test suffers from high false-negative rates and delayed results. Medical imaging, particularly Computed Tomography (CT) scans, has been recommended by the World Health Organization (WHO) as a complementary source of data for diagnosis and severity assessment of COVID-19. Countries with high rates of COVID-19 cases use chest CT images as the primary screening/monitoring technique. Several studies, therefore, have investigated COVID-19 manifestations on chest medical images. The most commonly observed chest imaging patterns in COVID-19 patients, as shown in Figure~\ref{fig:lunginf}, are Ground-Glass Opacity (GGO) and consolidation~\cite{ChungM:2020, ZhouSh:2020, ZhangX:2020}. GGO refers to a slight increase in lung attenuation such that the underlying vessels are still observable~\cite{Hansell.D.M.:2008}. The  consolidation, on the other hand, is considered as a rise in lung intensity such that the underlying vessels are obscured~\cite{Hansell.D.M.:2008}. The appearance of different types of imaging patterns and their location and distribution can be considered as specific signs of COVID-19 and provide helpful information for identifying the stage and severity of the disease~\cite{YuanM:2020, PanY:2020, SongF:2020, BernheimA:2020}.

Evaluation and quantification of lung involvement in COVID-19 patients based on their chest images can help determine the disease stage, have an optimal allocation of the limited health resources, and make informed treatment decisions. Compared to other imaging modalities, CT imaging provides more accurate representations of COVID-19 lesions, making it the most informative imaging modality for the prognosis of COVID-19 pneumonia. Radiologists measure the COVID-19 lesions from the chest CT images and quantify the disease's severity using different severity measures such as the Percentage of Opacity (PO) and CT severity score. The PO indicates the extent of involvement of the whole lung volume~\cite{chaganti2020automated}, while the CT severity score is determined based on the spread of the COVID-19 lesions in each lobe~\cite{francone2020chest}.  During the pandemic era, when the number of patients is exponentially increasing, visual assessment and quantification of lung lesions by expert radiologists become expensive, laborious, and prone to error. Automatic segmentation of infectious regions can, therefore, help quantify the extent of lung involvement  in patients confirmed with COVID-19, compute different severity scores, and speed up the treatment procedure.

\vspace{.025in}
\noindent
\textbf{Contributions:}
Motivated by the urgent quest to develop accurate and reliable automated models for prognostic assessment of COVID-19 pneumonia, we  introduce an open-access COVID-19 segmentation dataset along with a Deep Learning (DL)-based framework for the segmentation of COVID-19 lung abnormalities from chest CT scans. In summary, the main contributions of this study are as follows:
\begin{itemize}[noitemsep]
\item We propose the $\CR$ framework, which is a DL-based model for segmenting COVID-19 lesions, including GGOs and consolidations. In the $\CR$ architecture, a convolution layer with stride of two replaces the max-pooling layer to mitigate information loss during the down-sampling stage. Furthermore, we incorporate a context perception boosting  module in the encoding block that can learn the multi-scale representation of COVID-19 manifestations using four parallel paths of dilated convolutions and a linear projection of the input feature maps. Residual links in the encoding block help to learn in-depth features and avoid vanishing/exploding gradients. Implementing an un-symmetric network architecture reduces the model's complexity and computational time while slightly improving its  performance. Adopting a hybrid loss function facilitates image-level and patch-level  supervision during the training phase.
\item We produced a high-quality COVID-19 segmentation dataset containing $433$ annotated chest CT images from $82$ COVID-19 patients, annotated by a thoracic radiologist with $20$ years of experience.
\item A novel synthetic data generation (augmentation) pipeline  is proposed that generates synthetic pairs of CT images and infection masks by  inserting the infectious regions from COVID-19 CT images into healthy CT images. The proposed method improves the model performance by introducing more variability to the training set.
\end{itemize}
Based on a set of comprehensive experiments, we have evaluated performance of the proposed $\CR$ model in segmenting COVID-19 lesions from 2D CT slices and whole lung volumes. The above-mentioned contributions of the $\CR$ framework collectively have resulted in the state-of-the-art dice score of $80.2$\%, specificity of $99.7$\%, and sensitivity of  $83.2$\%. Additionally, experiments performed on the entire lung volumes indicate promising results, demonstrating that despite being trained only on infected CT images, the model can assist in patient-level lesion segmentation.

\section*{Methods} \label{sec:methods}
\begin{figure}[t!]
\centering
\includegraphics[width=1\linewidth]{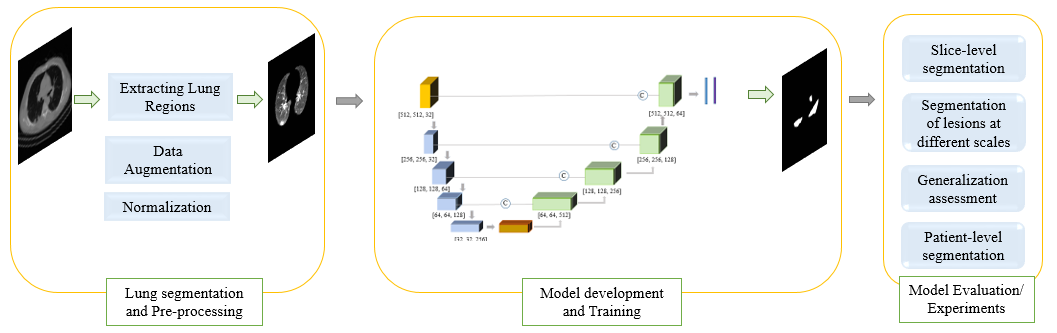}
\caption{\small The overall pipeline of the proposed $\CR$ framework.}
\label{fig:pipeline}
\end{figure}

In this section, we introduce the pixel-level labeled COVID-19 CT dataset along with the $\CR$ segmentation framework that takes thick-slice chest CT scans of confirmed COVID-19 patients and automatically segments regions of COVID-19 infection. Figure~ \ref{fig:pipeline} illustrates the overall pipeline of the $\CR$ framework, which consists of lung extraction and pre-processing stage, development and training stage,  and a set of comprehensive experiments for model evaluation. Additionally, a novel synthetic data generation (augmentation) pipeline is introduced to tackle the issue of limited access to annotated training data.

\subsection*{$\CRs$: Pixel-level Labeled COVID-19 CT Dataset}
As stated previously, DL-based segmentation networks need a large number of annotated CT images for efficient training. Most of the private annotated datasets, which are the basis of  the existing COVID-19 lesion segmentation models, are not  publicly available. Open-source pixel-level annotated datasets can promote developing DL networks for COVID-19 lesions segmentation. Furthermore,  evaluation of segmentation models on public datasets makes it feasible to fairly compare different segmentation models. To address the aforementioned issues, via this manuscript, we introduce a pixel-level annotated CT dataset, referred to as the $\CRs$. The CT scans used to generate the COVID-19 segmentation dataset are a subset of our previous COVID-19 diagnosis CT dataset~\cite{datasetDlink}, which is accessible through Figshare~\cite{datasetDlink}. A SIEMENS, SOMATOM Scope scanner has been used to obtain all axial CT images in $\CRs$. All CT images are reconstructed by the Filtered Back Projection method using the reconstruction matrix size is $512 \times 512$ and D40s kernel, which modifies the data frequency contents and mitigates the noise. All CT images are obtained without contrast enhancement and saved in the Digital Imaging and Communications in Medicine (DICOM) format and the Hounsfield Unit~\cite{datasetDlink}.

\begin{figure}[t!]
\centering
\includegraphics[width=0.9\linewidth]{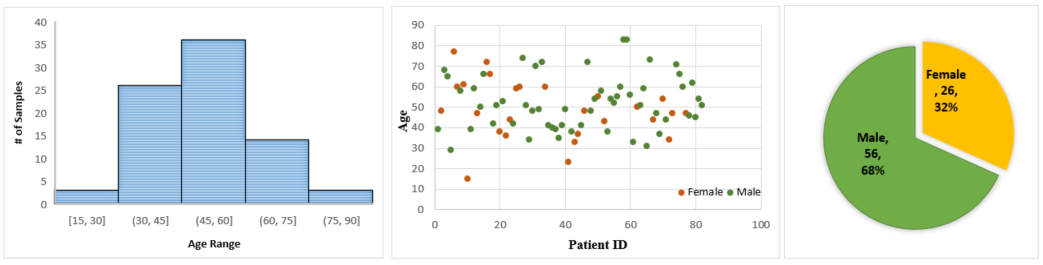}
\caption{\small Age and gender distribution of the  $\CRs$ dataset.}
\label{fig:datasetStat}
\end{figure}

\begin{figure}[t!]
\centering
\includegraphics[width=0.9\linewidth]{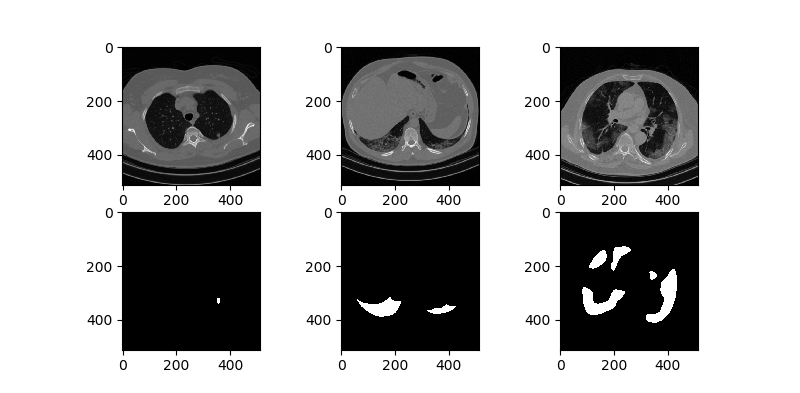}
\caption{\small Examples of CT images with various infection rates and different positions in lung volume. First row: CT images. Second row: Ground truth masks.}
\label{fig:in-house-data}
\end{figure}

In the annotation process, a standard U-Net model is firstly trained on an open-access COVID-19 segmentation dataset~\cite{datasetA}. The trained model then takes the CT images as unseen test sets and predicts the infection masks. Next, a thoracic radiologist with $20$ years of experience in lung imaging, carefully modified and verified the generated infection masks. Overall, we annotated $433$ CT images from $82$ COVID-19 patients. The patients' average age is $51 \pm 13.59$ years (mean $\pm$ std). Figure~\ref{fig:datasetStat} represents the age and gender distribution of the COVID-19 patients in the $\CRs$. We annotated GGOs and consolidations as the most prevalent manifestations of COVID-19. The CT images from diffident parts of the lung (top, middle, and bottom) are  included in the dataset with different infection rates (area of the infection divided by the lung area in each CT image). These considerations help the AI models to achieve  better performance when predicting the regions of infection on a whole lung volume. Examples of CT images with their ground truth masks are shown in Figure~\ref{fig:in-house-data}.

\subsection*{COVID-19 Lesion Segmentation Network}
\begin{figure}[t!]
\centering
\includegraphics[width=0.8\linewidth]{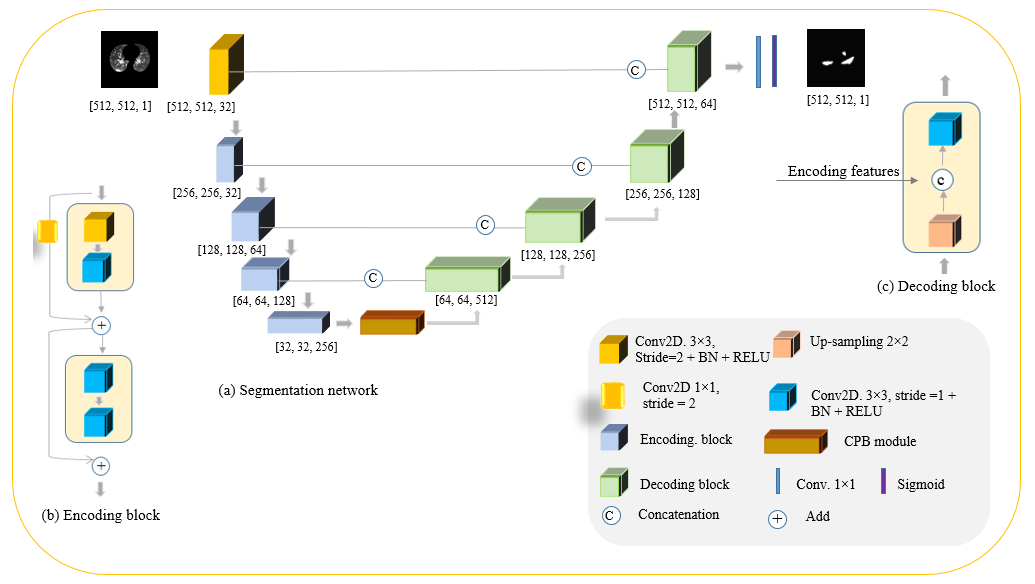}
\caption{\small Architecture of the proposed $\CR$ segmentation framework.}
\label{fig:segnet}
\end{figure}

The architecture of the proposed $\CR$ framework (as shown in Figure~\ref{fig:segnet}) is an encoder-decoder-based network containing an encoding path, a context perception boosting module, and a decoding path. These three underlying components are described below:

\vspace{.025in}
\noindent
\textit{Encoding Path} extracts informative features from chest images. The Encoding Path is initiated with a $3 \times 3$ convolution layer with $32$ filters, followed by a Batch Normalization (BN) layer and ReLU activation function. Four encoding blocks with $32$, $64$, $128$, and $256$ filters, with the architecture represented in Figure~\ref{fig:segnet}(b), are then applied sequentially. The encoding block has two units, each consisting of two successive $3 \times 3$ convolution layers, followed by the BN layer and ReLU function. The first convolution layer of the first unit in the encoding block uses a stride of $2$ for down-sampling. The max-pooling layer is replaced with a convolution layer with a stride of two to mitigate information loss during down-sampling. The input feature maps of each unit are then added to the  output feature maps using element-wise addition operation. Transmitting the feature maps to the deeper levels (residual linking) facilitates the convergence of the network and alleviates the gradient vanishing issue~\cite{resnet2016}.

\begin{figure}[t!]
\centering
\includegraphics[width=0.4\linewidth]{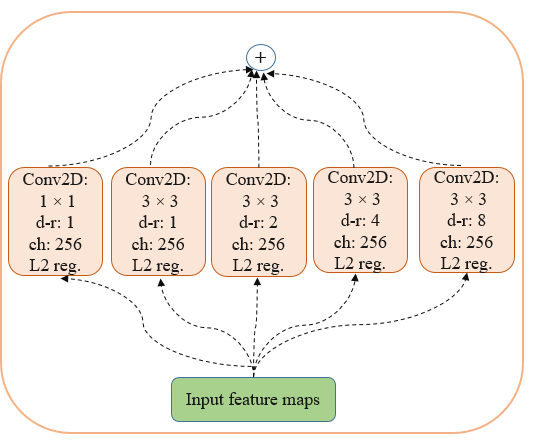}
\caption{\small Architecture of the proposed context perception boosting module.}
\label{fig:cpb}
\end{figure}

\vspace{.025in}
\noindent
\textit{Context Perception Boosting (CPB) Module:} As mentioned previously, a critical challenge in segmenting COVID-19 lung abnormalities is various sizes and scattered distribution of the  regions of infection. Since a Convolutional Neural Networks (CNN)-based encoder extracts the information from a local area of the global image by applying the convolution kernel, it may fail to learn the long-range dependencies of the global image. To address this problem and inspired by~\cite{ChenL2017}, we adopt a context perception boosting module in the last encoding block of the $\CR$ network. This module, as demonstrated in Figure~\ref{fig:cpb}, contains a $1 \times 1$ convolution layer for linear projection of the input feature maps and four parallel $3 \times 3$ convolution layers with different dilation rates. $L2$  regularization is applied after each convolution operation to penalize weight matrices and mitigate over-fitting. The outputs of different convolution layers are then integrated using element-wise addition. The dilation rates for parallel convolutions are set to $1$, $2$, $4$, and $8$. The context perception boosting module increases the receptive field of the extracted feature maps by incorporating multi-scale kernels, helping the network detect COVID-19 abnormalities of various sizes.

\vspace{.025in}
\noindent
\textit{Decoding Path}, takes the extracted features of the context perception boosting  module as input, aiming to restore the spatial representation and generate masks indicating the regions of infection. For reconstructing precise infection masks, four decoding blocks are adopted within the decoder path of the $\CR$ segmentation network. Each decoding block, presented in Figure~\ref{fig:segnet}(c), starts with an up-sampling layer to expand the feature maps' size to the size of encoding feature maps. The up-sampling layer's outputs are then concatenated to the extracted features from the encoding blocks. A $3 \times 3$ convolution layer followed by a BN layer, and a ReLU layer is then applied to the skip connection's output to adjust the number of feature channels and learn the spatial representation of the contextual information. The number of filters for decoding blocks are $512$, $256$, $128$, and $64$. The first decoding block of the $\CR$ network has the maximum number of filters. Indeed, the number of feature channels in each decoding block is not equal to its corresponding encoding block, leading to fewer learning parameters, reducing the model complexity, and saving computational time. The output of the last decoding block has the same spatial dimension as the original CT scans. A $1 \times 1$ convolution layer with one feature channel is applied on the last decoding block's output. The sigmoid activation function predicts the probability of each pixel belonging to the infection class.

\subsection*{Synthetic Data Generation Specific to COVID-19 Lesion Segmentation}
Using data augmentation techniques~\cite{RonnebergerO:2015, OliveiraA2017} has become a standard solution in different computer vision tasks to tackle the lack of sufficient training data and prevent over-fitting. These techniques mainly include applying simple transformations such as rotating, horizontal/vertical flip, zooming, and/or translation on original images. However, these transformations make minor changes to the existing images, resulting in limited diversity on newly generated images. Development of new data augmentation techniques are therefore crucial to introduce more variability to the synthetic samples for better data reinforcement. For example, recently, Reference~\cite{XuZ2020} implemented a specific data augmentation approach into a generative adversarial segmentation network and achieved improved results. Capitalizing on this vision, we propose a specific approach for generating synthetic pairs of CT images and their corresponding infection masks by extracting the COVID-19  regions of infection from infected chest CT scans and inserting them into healthy chest CT images. We use a chest CT image containing COVID-19 regions of infection alongside its infection mask and a chest CT image from a healthy person acquired by the same scanner alongside its lung mask as the input. First, CT images are normalized using the mean and standard deviation of their pixel intensities. Next, regions of infection  are extracted from the COVID-19-infected CT image using the infection mask. Lung regions from the normalized healthy CT image are extracted using its corresponding lung mask. Then, the location of the region of infection  is evacuated by element-wise multiplication of normalized healthy lung regions and the inverted infection mask. In the next step, the regions of infection  are added to the evacuated healthy CT image. Finally, the regions of infection  outside the healthy lung regions are trimmed off using the healthy lung mask, and the new infected CT image is generated. The synthetic infection mask is adjusted for the synthetic infected CT image by multiplying the healthy lung mask by the infection mask.

\subsection*{Hybrid Loss Function}
For a segmentation task, when dealing with highly imbalanced datasets, which is common when dealing with COVID-19 chest CT scans, equally penalizing False Negatives (FNs) and False Positives (FPs) during the training process will result in a low sensitivity in predicted masks. To tackle this problem, we trained the $\CR$ model by minimizing a hybrid loss function over the training epochs to supervise the model in both image-level and local areas.  The hybrid loss is a combination of weighted Binary Cross-Entropy (w-BCE)~\cite{afshar20203d} and Focal Tversky Loss (FTL)~\cite{abraham2019novel} and is given by
\begin{equation}
\textit{$L_{Total}$}=\textit{$L_{w-BCE}$}~~~+~\kappa~~\textit{$L_{TFL}$},
\end{equation}
where $\kappa$ determines the FTL's contribution to the total loss function and is set to 1 in our experiments. The FTL, which is a generalization of the Tversky Index (TI)~\cite{salehi2017tversky}, is a loss function that improves the model performance by assigning higher weights to hard pixels and has been introduced explicitly for the segmentation of class-imbalanced datasets. The TI and FTL are defined as follows
\begin{eqnarray}
\text{$TI$} &=& \frac{{\sum_{i=1}^{N}}{P_{li}~GT_{li}}}{{\sum_{i=1}^{N}}{P_{li}~GT_{li}}~~+~~\alpha~~ {\sum_{i=1}^{N}}{P_{li}~GT_{bi}}~~+~~\beta~~{\sum_{i=1}^{N}}{P_{bi}~GT_{li}}}\\
\text{$FTL$} &=& \sum_c{(1 - TI)}^\frac{1}{\gamma}
\end{eqnarray}
where $P_{li}$ is the probability that pixel $i$ belongs to the lesion class and $P_{bi}$ is the probability that pixel $i$ is of the background class. The $GT_li$ is one for a pixel labeled as lesion and zero for a pixel labeled as background, and vice versa for the $GT_bi$. The trade-off between FPs and FNs can be adjusted through the hyper-parameters $\alpha$ and $\beta$. Term $\gamma$ is defined within the range of [$1,3$] and forces the model to detect small Regions of Interest (RoI). We set $\alpha = 0.7$, $\beta = 0.3$, and $\gamma = \frac{4}{3}$~\cite{abraham2019novel}.  Compared to the standard $L_{BCE}$, the $L_{w-BCE}$ can improve the model's performance by assigning higher weights for the target pixels (i.e., COVID-19 lesions) instead of weighting all pixels equally. 

\subsection*{Experiments}
This research work is performed based on Concordia University's ethics clearance with the policy certification number of $30013394$. To evaluate the efficacy and potential limitations of the proposed $\CR$ framework, we performed an extensive set of experiments. In particular, we have investigated the model's capability in detecting regions of infection of different sizes  on the CT scans. The conducted experiments analyze the model's performance in segmenting COVID-19 lesions in: (i) Slice-level segmentation, where only CT images containing COVID-19 abnormalities are fed to the network, and; (ii) Patient-level experiments when the model is tested on the whole CT volume containing CT images with and without COVID-19 lesions.  To explore $\CR$'s generalization, we experimented on an independent dataset and evaluated the model's performance on CT images acquired by a different scanner in a different imaging center. We also tested the $\CR$ on CT volumes of $50$ COVID-19 patients to assess performance of the model in discriminating infected CT images from non-infected ones. 

\subsubsection*{Description of Datasets}
Four datasets are used in different steps of the experiments as outlined below:
\begin{itemize}
\item \textbf{\textit{Dataset A}}~\cite{datasetA}, which is a public dataset containing chest CT volumes of $10$ COVID-19 patients, a total of $2,581$ CT slices. The matrix size of the images is $512\times512$ pixels. In total, $1,351$ CT slices (out of $2,581$) have shown COVID-19 manifestations and have been annotated by three expert radiologists. The segmentation of lung regions exists for all CT slices.
\item \textbf{\textit{Dataset B (the $\CRs$ dataset)}:} The in-house dataset, which contains $433$ annotated CT images from $82$ COVID-19 patients.
\item \textbf{\textit{Dataset C}}~\cite{datasetC}, which contains nine COVID-19 chest CT volumes, a total of $829$ CT slices. In total, $373$ out of $829$ CT images indicated COVID-19 lesions and have been segmented by a radiologist. The segmentation of lung regions has been performed for the whole CT volumes. The CT slices have a dimension of $630\times630$ pixels and are resized to $512\times512$ pixels for our experiments.
\item \textbf{\textit{Dataset D}}, introduced in our previous work \cite{datasetDlink}, containing chest CT volumes from $76$ healthy cases, $171$ COVID-19 patients, and $60$ patients with other types of Community Acquire Pneumonia (CAP). In this dataset, $55$ out of $171$ COVID-19 CT volumes have slice-level labels, indicating which CT slices demonstrate COVID-19 lesions. The CT images used for the $\CRs$ dataset are a subset of COVID-19 cases in Dataset D.
\end{itemize}
The first two datasets (Dataset A and B) are used for training/testing purposes, while the third and fourth datasets (Dataset C and D) are used to evaluate the model's performance from various aspects.
\subsubsection*{Evaluation Metrics}
We evaluate the $\CR$ model's performance in segmenting COVID-19 lesions by comparing the predicted segmentation masks with the ground truth labels. The following  metrics are used for evaluation purposes:

\vspace{.05in}
\noindent
\textbf{\textit{Dice Similarity Coefficient (DSC)}}, which is the most commonly used metric for segmentation. The DCS measures the relative overlap of the predicted regions of infection  and their ground truth labels. It takes the value one as its maximum when the two regions have a complete agreement and a minimum value of zero when there is no overlap between the two underlying areas. The DSC is defined as follows
\begin{eqnarray}
\text{DSC} = \frac{2 (|Pr| \cap |GT|)}{|Pr| + |GT|},
\end{eqnarray}
where terms $Pr$ and $GT$ correspond to the set of pixels belonging to the predicted and the ground truth regions, respectively. The symbol $\cap$ represents the intersection operation, and $|\cdot|$ is the cardinality operator.

\vspace{.05in}
\noindent
\textbf{\textit{Sensitivity (SEN) and Specificity (SPC)}}: The SEN metric calculates the number of pixels correctly labeled as COVID-19 lesions in the predicted masks relative to the total number of pixels identified as COVID-19 abnormalities in the ground truth. The SPC metric measures the number of pixels correctly labeled as the background class relative to the total number of background pixels in the ground truth. SEN and SPC metrics are defined as follows
\begin{eqnarray}
SEN &=& \frac{TP}{TP + FN},\\
\text{and } \quad SPC &=& \frac{TN}{TN + FP},
\end{eqnarray}
where $TP$, $FN$, $TN$, and $FP$ are the number of pixels in the true positive, false negative, true negative, and false-positive regions, respectively.

\vspace{.05in}
\noindent
\textbf{\textit{Mean Absolute Error (MAE)}}, which calculates the absolute error between each pixel's predicted and ground-truth label and takes the average over the whole pixels. The MAE is given by
\begin{equation}
MAE = \frac{1}{w \times h}{\sum_{j=1}^{w}} {\sum_{i=1}^{h}} \text{ \big|$Pr_{ij}$ - $GT_{ij}$\big|}.
\end{equation}
The MAE metric has a minimum of zero if all the pixels are correctly labeled and a maximum of one when all the pixels are predicted with the wrong label.

\subsubsection*{Pre-processing Step}
In the pre-processing step, we extract lung regions for all CT images using a lung segmentation model, referred to as the ``U-net (R231CovidWeb)"~\cite{HofmanningerJ:2020}. This model has been trained and evaluated on three public datasets covering six different types of lung diseases and then fine-tuned on a small COVID-19 dataset. After segmenting the lung regions, only CT images containing the lung tissues are passed to the next stage, and the rest are eliminated. Each CT image is normalized based on its mean and standard deviation (std). The datasets are divided into three independent groups for training ($60$\%), validation ($10$\%), and testing ($30$\%). To avoid information leakage, we kept the underlying datasets patient-independent, meaning that patients' CT images are not shared between training, validation, and test sets. Furthermore, to improve the model performance on unseen data and mitigate over-fitting issues, we use real-time data augmentation strategies, including zooming, shifting, and shearing, where artificial images are synthesized from each mini-batch of original CTs during the training process. The model observes each synthetic image only once, resulting in an enhancement in the model's overall generalization ability. The network uses the Adam optimizer with an initial learning rate of $0.001$ over $100$ epochs and will stop if the loss function on the validation set does not decrease over ten epochs.

\subsubsection*{Quantitative Analysis}
\begin{table}[t!]
\centering
\begin{small}
\caption{\small Performance of the proposed segmentation network on the test set through 10-fold cross-validation approach}
\label{tab:res10-fold}
\begin{tabular}[c]{c c c c c}\\
\hline
 & \textbf{DSC} & \textbf{SPC} & \textbf{SEN} & \textbf{MAE}\\
\hline
\textbf{$Ave \pm std$} & $0.8019 \pm 0.032$ & $0.9969 \pm 0.0008$ & $0.8323 \pm 0.046$ & $0.0052 \pm 0.001$\\
\hline
\end{tabular}
\end{small}
\end{table}
\begin{table}[t!]
\centering
\begin{small}
\caption{\small Quantitative evaluation of our proposed model trained with different loss functions. The presented results are the average of the obtained results through a 10-fold cross-validation process. The best results have been highlighted in bold.}
\label{tab:lossfunc}
\begin{tabular}[c]{c c c c c}\\
\hline
\textbf{Loss Function} & \textbf{DSC} & \textbf{SPC} & \textbf{SEN} & \textbf{MAE}\\
\hline
w-BCE & 0.7806 & 0.9950 & \textbf{0.8399} & 0.0058\\
FTL & 0.7919 & 0.9967 & 0.8134 & 0.0054\\
Hybrid loss function & \textbf{0.8019} & \textbf{0.9969} & 0.8323 & \textbf{0.0052}\\
\hline
\end{tabular}
\end{small}
\end{table}
We use a combination of Datasets A and B, i.e.,  a total of $1,784$ CT images with COVID-19 manifestations from $92$ patients to train/test the proposed $\CR$ model. The average of the evaluation metrics calculated on the whole test set over $10$-fold cross-validation is represented in Table~\ref{tab:res10-fold}. The average DSC over the entire test set over $10$-fold cross-validation is equal to $0.8019$. The average MAE is  $0.0052$, which means that $99.48$\% of pixels on the test set have been labeled correctly. The average of SPC and SEN metrics are $0.9969$ and $0.8323$, respectively. The calculated standard deviations for the evaluation metrics demonstrate that the different folds' results do not vary significantly, indicating that the model has a reliable performance on different test sets of the $10$-fold cross-validation.

We assess effects of different loss functions, including weighted binary cross-entropy (w-BCE), FTL, and the hybrid loss function on the model performance; results are shown in Table~\ref{tab:lossfunc}. The FTL loss function improves the model performance (DCS, SPC, and MAE metrics) compared to the w-BCE loss function. However, it decreases the model sensitivity from $83.99$\% to $81.34$\%. The model trained on the hybrid loss function slightly reduces the SEN  compared to the w-BCE loss but improves DSC, SPC, and MAE evaluation metrics.

\begin{figure}[t!]
\centering
\includegraphics[width=0.8\linewidth]{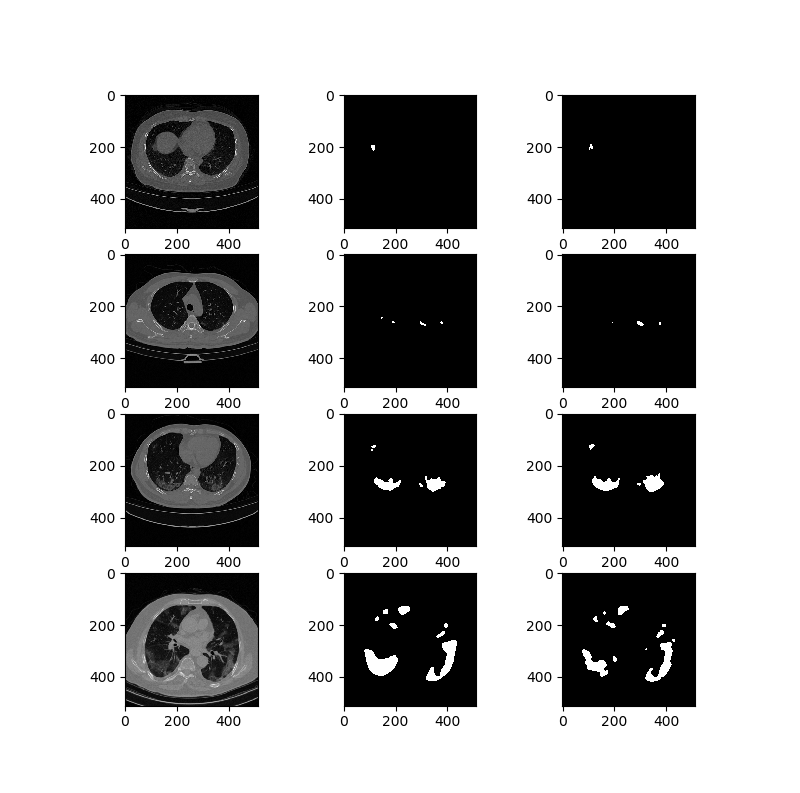}
\caption{\small Qualitative evaluation of model performance on test set. From left: Original image, Ground truth mask, Predicted mask. Row 1 and 2: Test images from group A with infection rate less than 0.015, Row 3 and 4; Test images from group B with infection rate more significant that 0.015.}
\label{fig:goodRes}
\end{figure}

To better investigate our model's performance in segmenting COVID-19 lesions at different scales, we categorize the test images into two groups: (i) \textit{Group-A}, CT images with infection rate smaller than $0.015$, and; (ii) \textit{Group-B}, CT images with the infection rate larger than $0.015$. The infection rate is calculated by dividing COVID-19 lesions' area by the lungs' area in a CT slice. Table~\ref{tab:groupAB} represents performance of the $\CR$ model across these two groups, indicating the median of evaluation metrics and their Inter-quartile Ranges (IQR) for $25\%$ and $75\%$ percentiles. As can be seen, although the model yields better results on CT images in Group-B, its performance in dealing with small regions of infection is also acceptable. The average DCS for Group-A (median: 0.6748, IQR: 0.3035-0.8007) and Group-B (median: 0.8254, IQR: 0.7379-0.8751) are calculated. The quantitative analysis over Group-A and Group-B indicates that the proposed model is reliable in segmenting COVID-19 lesions at different scales. To better understand the model's capabilities, we visualized some examples of Group-A and Group-B along with their predicted masks in Figure~\ref{fig:goodRes}.

For comparative analysis, we compare the proposed $\CR$ framework with the following benchmark models: (i) Standard U-Net; (ii) Residual U-Net with CPB module, where the number of feature channels starts from $32$ in the first residual block in the contracting path and reaches $512$ in the bottleneck, and; (iii) $\CR$ without CPB module. Table~\ref{tab:architec} presents the average of different evaluation metrics together with the number of trainable parameters of each network architecture.  A common 10-fold cross-validation with similar data split is used to train/test different models.

\begin{table}[t!]
\centering
\begin{small}
\caption{\small Performance of the proposed $\CR$ network on CT images of groups A and B. The presented results are the average of the 10-fold cross-validation.}
\label{tab:groupAB}
\begin{tabular}[c]{c c c}\\
\hline
 & \textbf{Group A} & \textbf{Group B}\\
 & median (IQR 25\%, IQR 75\%) & median (IQR 25\%, IQR 75\%)\\
\hline
\textbf{DSC} & 0.6748 (0.3035, 0.8007) & 0.8254 (0.7379, 0.8751)\\
\textbf{SPC} & 0.9997 (0.9990, 0.9999) & 0.9981 (0.9961, 0.9992)\\
\textbf{SEN} & 0.7419 (0.3283, 0.9240) & 0.8909 (0.7522, 0.9521)\\
\hline
\end{tabular}
\end{small}
\end{table}

\begin{table}[t!]
\centering
\begin{small}
\caption{\small Quantitative comparison of different architectures in segmenting COVID-19 lesions. The presented results are the average of the obtained results through a 10-fold cross-validation process. The best results have been highlighted in bold.}
\label{tab:architec}
\begin{tabular}[c]{c c c c c c}\\
\hline
\textbf{Architecture} & \textbf{Trainable parameters (m)} & \textbf{DSC} & \textbf{SPC} & \textbf{SEN} & \textbf{MAE}\\
\hline
Standard U-Net & 7.85 & 0.7793 & 0.9963 & 0.7622 & 0.0059\\
Residual U-Net with CPB & 20.3 & 0.7921 & 0.9968 & 0.8239 & 0.0055\\
$\CR$ without CPB & 6.32 & 0.7959 & \textbf{0.9969} & 0.8283 & 0.0054\\
$\CR$ with CPB & 8.75 & \textbf{0.8019} & \textbf{0.9969} & \textbf{0.8323} & \textbf{0.0052}\\
\hline
\end{tabular}
\end{small}
\end{table}

\subsubsection*{Generalization: Assessment on External Datasets}
When developing AI-based models using CT scans, it is crucial to assess if the trained model can be generalized on CT images from a new scanner. This is mainly because CT scans, acquired from different scanners with varying acquisition settings, show different resolutions and characteristics. For this purpose, we evaluate the $\CR$ performance on a third dataset (referred to as Dataset C) that is different from our training set and contains COVID-19 CT images from nine patients. We ignore one of the CT volumes that shows minimal infection regions (the infection rate is almost zero) and include the rest in the experiments. In the first step of this generalization assessment experiment, we examine the $\CR$ model only on CT images containing COVID-19 lesions ($372$ out of $787$ CT scans contained infection regions). The results are shown in Table~\ref{tab:externaldata}. The DSC, SPC, SEN, and MAE metrics for the first test are $0.797$, $0.9933$, $0.898$, and $0.0085$. The obtained results show that the model can reasonably work on CT images from a new scanner. We compare performance of the $\CR$ with Reference~\cite{ZhouT:2020} that has used $374$ infected CT slices from Dataset C along with $100$ infected CT slices for training and testing their proposed segmentation network through a 5-fold cross-validation. Although their model has been trained on this dataset, the proposed $\CR$ framework outperforms their model in terms of SPC and SEN metrics and yields comparable DCS results.

\begin{figure}[t!]
\centering
\includegraphics[width=0.65\linewidth]{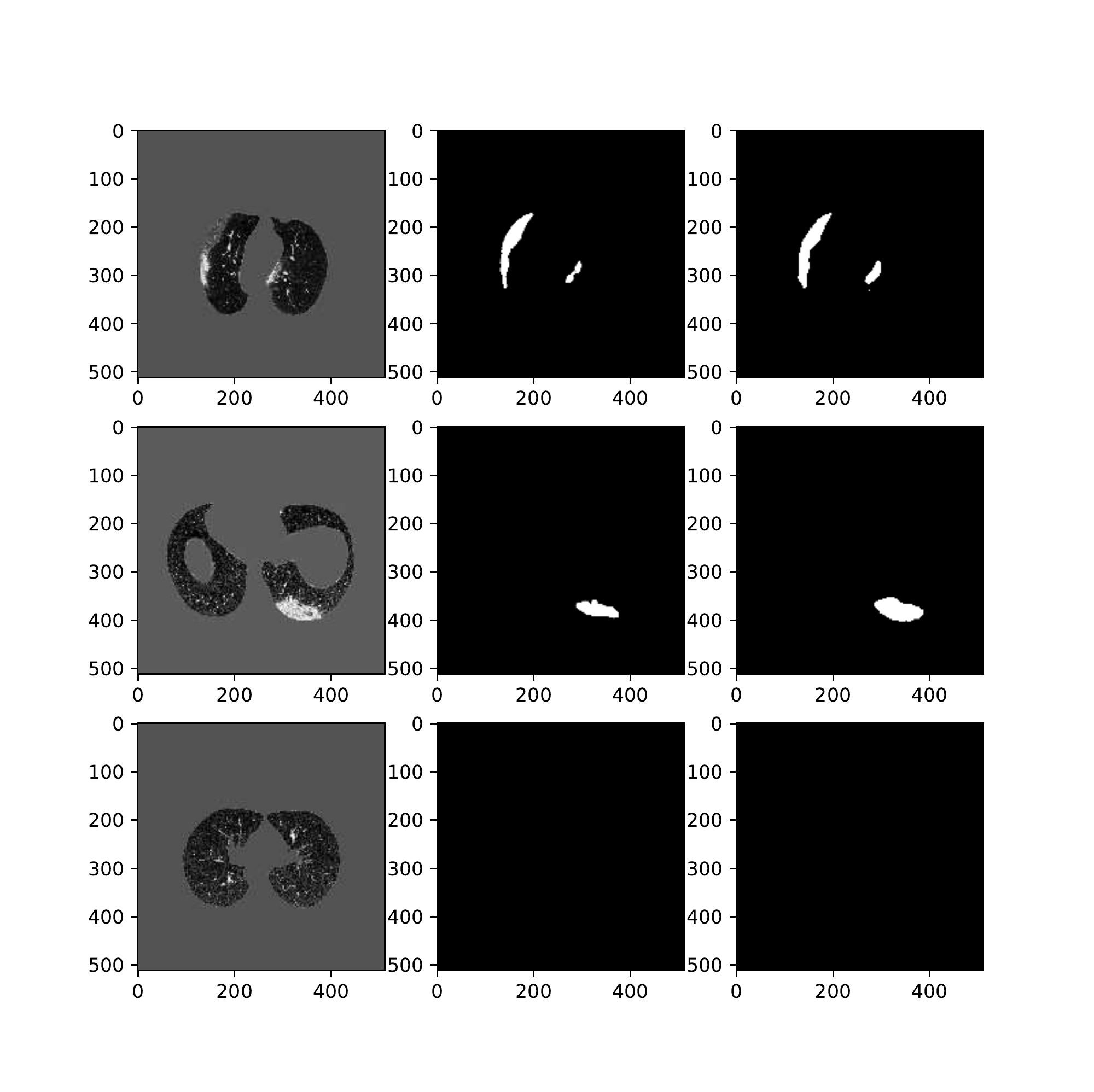}
\caption{\small Qualitative evaluation of the model generalization on an external dataset. From left: Original image, Ground truth mask, Predicted mask. The model predicts a black mask for CT images with no trace of viral infection.}
\label{fig:gnz_plt}
\end{figure}
\begin{figure}[t!]
\centering
\includegraphics[width=0.4\linewidth]{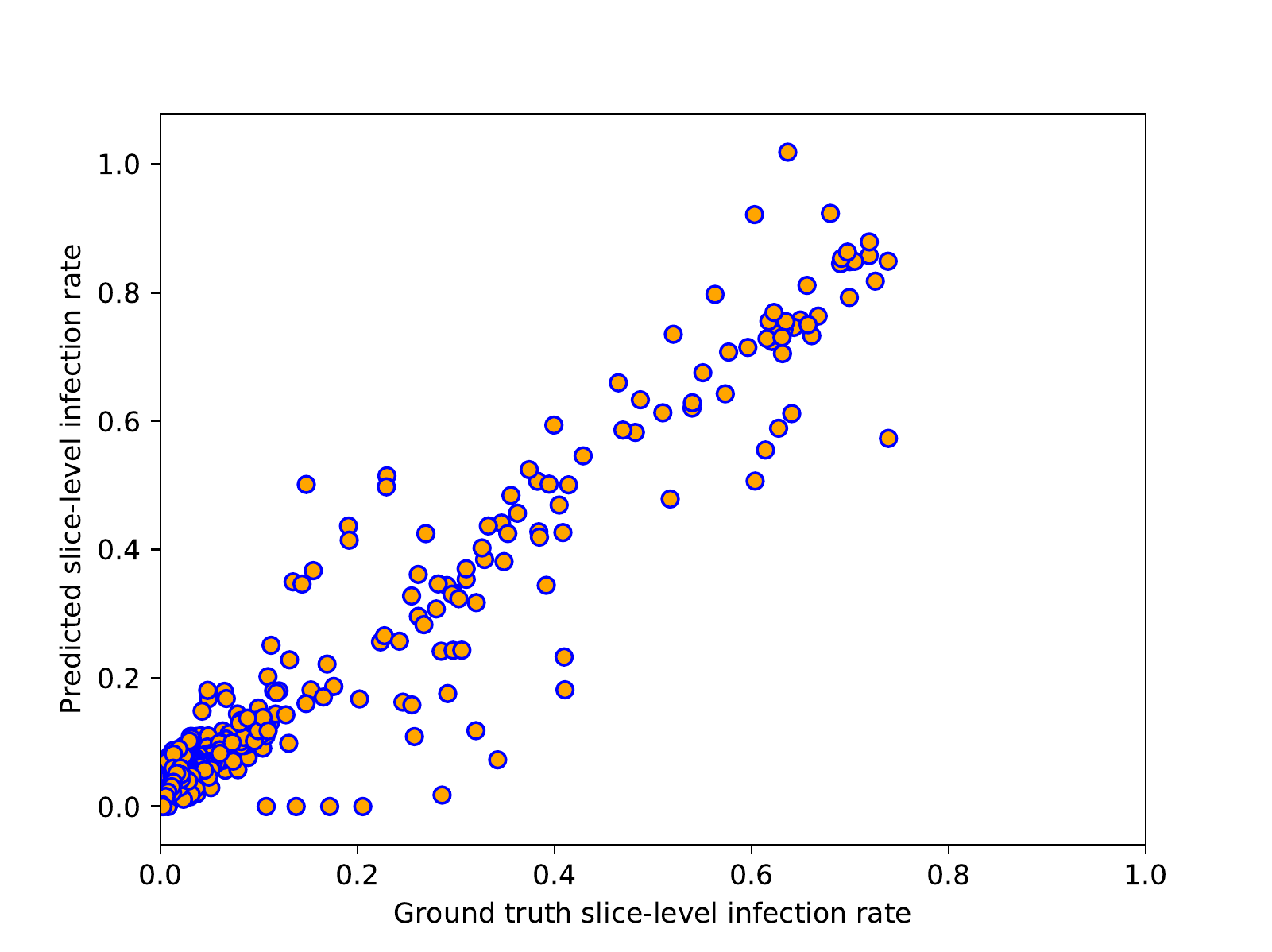}
\caption{\small Ground truth and predicted slice-level infection rate in generalization assessment experiment.}
\label{fig:infrate}
\end{figure}

In practical settings, COVID-19 CT volumes consist of several slices, some of them with regions of infection  and the rest with no evidence of infection. To assess the model's performance in segmenting COVID-19 lesions, we test the $\CR$ on whole lung volumes of eight COVID-19 patients ($683$ out of $787$ CT scans included lung tissues). Figure~\ref{fig:gnz_plt} demonstrates some examples of the ground truth and the predicated masks from this experiment. As can be observed, the model predicts a black mask for CT images with no trace of infection. The experiment results over the $683$ test images containing lung tissues are $0.79$, $0.996$, $0.893$, and $0.0048$ for DSC, SPC, SEN, and MAE metrics. Compared to Reference~\cite{fan2020inf}, which have used $638$ slices of this dataset ($285$ normal slices and $353$ slices with COVID-19 lesions) as an external test set for evaluation of their semi-supervised segmentation framework, the $\CR$ achieved better results across all the evaluation metrics. The mean absolute error between the ground truth and predicted infection rate for each CT image is $0.053$ and $0.008$ for CT scans with and without evidence of infection, respectively. Figure~\ref{fig:infrate} represents the linear relationship between ground truth and predicted infection rates for CT images containing COVID-19 lesions. The Pearson correlation coefficient between the two groups of data is measured $0.957$, indicating that the predicted and ground truth infection rates are highly correlated.

\begin{table}[t!]
\centering
\begin{small}
\caption{\small Model evaluation on external dataset on CT images containing COVID-19 lesions and on whole CT volumes.}
\label{tab:externaldata}
\begin{tabular}[c]{c c c c c c c}\\
\hline
\textbf{Method} & \textbf{Input data} & \textbf{Validation type} & \textbf{DSC} & \textbf{SPC} & \textbf{SEN} & \textbf{MAE}\\
\hline
Ref.~\cite{ZhouT:2020} & CTs with infection regions & cross-validation & \textbf{0.831} & 0.993 & 0.867 & -\\
$\CR$ & CTs with infection regions & External validation & 0.797 & \textbf{0.9933} & \textbf{0.898} & 0.0085\\
\hdashline
Ref.~\cite{fan2020inf} & whole lung volumes & External validation & 0.597 & 0.977 & 0.865 & 0.033\\
$\CR$ & whole lung volumes & External validation & \textbf{0.79} & \textbf{0.996} & \textbf{0.893} & \textbf{0.0048}\\
\hline
\end{tabular}
\end{small}
\end{table}

\subsubsection*{Model Evaluation on Discriminating CT Images with Infection}
To further explore how the proposed $\CR$ framework performs on whole CT volumes, we use a subset of Dataset D that contains $50$ CT volumes with slice-level labels and evaluate the $\CR$ performance on discriminating infected CT images from non-infected ones. According to our framework, we first extracted lung regions from CT images. From $6,846$ CT images containing lung tissues, $3,503$ showed COVID-19 lesions. Since our $\CRs$ dataset is a subset of Dataset D, we eliminated the $267$ common CT slices that were overlapping between the two datasets. The rest of the CT scans, a total of $6,579$ images from $50$ patients, were passed through the segmentation network. The model predicts a mask containing lesion regions for CT images with infection and a black mask for CT images without infection. To avoid spurious effects from minor imaging findings, we consider a CT scan as an ``infected image'' only if its predicted infection rate is more significant than $0.005$. The accuracy, sensitivity, and positive predictive value for this discrimination experiment are $0.871$, $0.94$, and $0.833$, which are promising for the potential application of the $\CR$ framework on whole lung volumes.

\subsubsection*{Efficacy of the Proposed Synthetic Data Augmentation}
\begin{figure}[t!]
\centering
\includegraphics[width=0.8\linewidth]{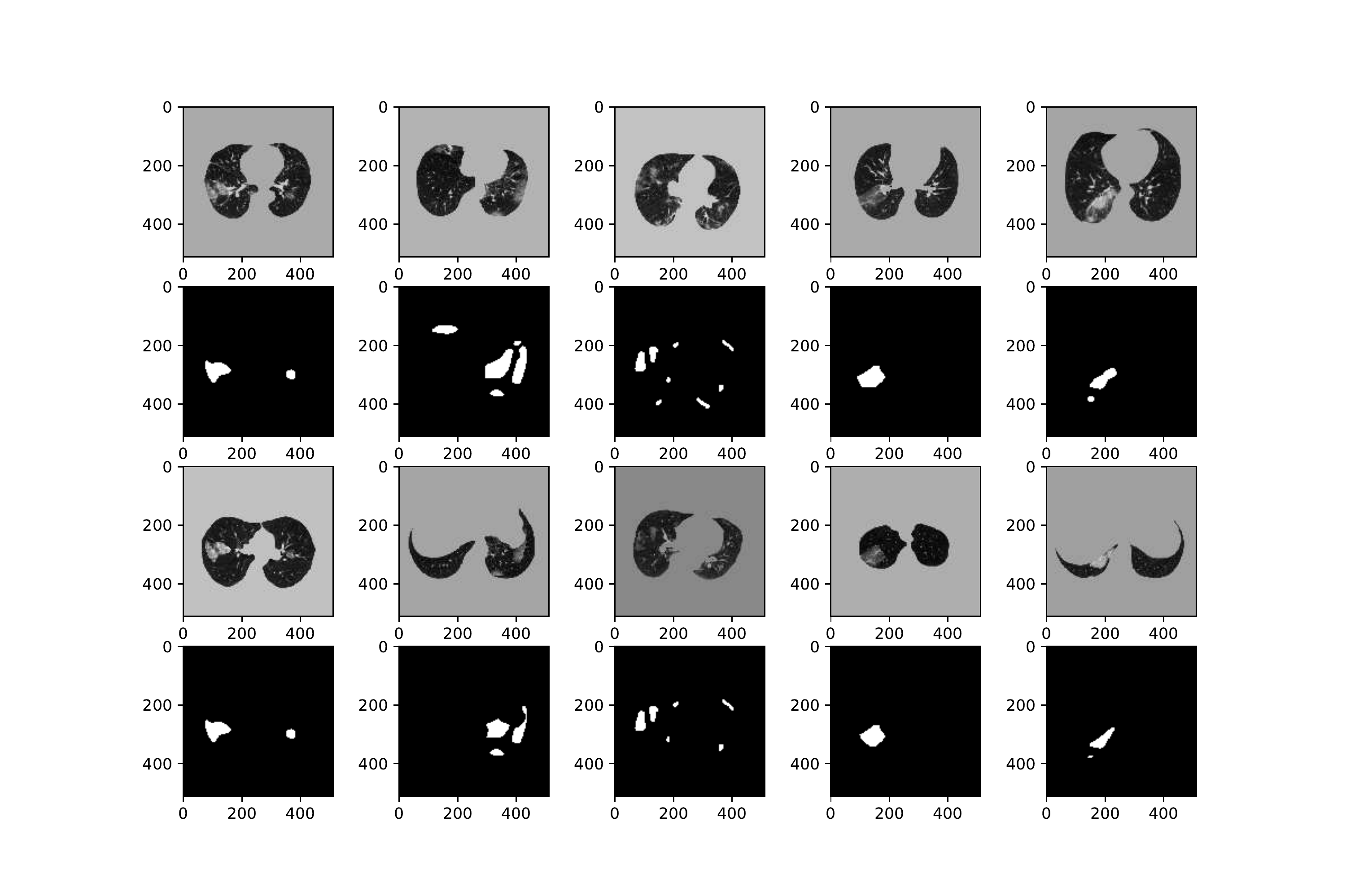}
\caption{\small Samples of synthetic images and infection masks. First and second rows: COVID-19 infected CT images and their corresponding infection masks. Third and fourth rows: Synthetic images and the adjusted infection masks.}
\label{fig:synth_im}
\end{figure}

To explore the efficacy of our proposed data augmentation technique in the training process, first, we trained the $\CR$ model without the CPB module on the COVID-CT-Rate dataset. We split the dataset into three subsets for training ($60$\%), validation ($10$\%), and testing ($30$\%). The conventional data augmentation techniques, including zooming, shifting, and shearing, were used during the training. In the next step, we performed the same experiment using a training set augmented by our data augmentation technique. For this purpose, we used $988$ CT images from nine healthy lung volumes acquired by the same scanner of the COVID-CT-Rate. In each fold of the cross-validation, we randomly selected a subset of healthy CT images with the size of the training set and generated synthetic pairs of CT images and infection masks based on the training set of each fold. Figure~\ref{fig:synth_im} demonstrates some samples of the synthetic images and their corresponding infection masks. The synthetic images with the infection rate more significant than $0.01$ are concatenated with the training set. In each fold, the validation and test sets are kept the same as the previous step of the experiment. The results of both experiments through 10-fold cross-validation are presented in Table~\ref{tab:res-aug}.

\begin{table}[t!]
\centering
\begin{small}
\caption{\small Evaluation of the proposed data augmentation method's efficacy through a two-step 10-fold cross-validation approach.}
\label{tab:res-aug}
\begin{tabular}[c]{c c c c c}\\
\hline
\textbf{Training set} & \textbf{DSC} & \textbf{SPC} & \textbf{SEN} & \textbf{MAE}\\
 & \textbf{$Ave \pm std$} & \textbf{$Ave \pm std$} & \textbf{$Ave \pm std$} & \textbf{$Ave \pm std$}\\
\hline
\textbf{Existing Training Set} & $0.7599 \pm 0.05$ & $0.9953 \pm 0.003$ & $0.8104 \pm 0.06$ & $0.0074 \pm 0.002$\\
\textbf{Augmented Training Set} & $0.781 \pm 0.04$ & $0.9957 \pm 0.002$ & $0.8308 \pm 0.03$ & $0.0067 \pm 0.001$\\
\hline
\end{tabular}
\end{small}
\end{table}

\section*{Discussion}
This study is motivated by the urgent quest to develop accurate and reliable automated models for prognostic assessment of COVID-19 pneumonia coupled with the need for providing high-quality COVID-19 lesion segmentation datasets. To better position this study within the existing literature, first, we discuss the related works.

\vspace{.025in}
\noindent
\textbf{Related Works:}
Recently, several autonomous models are proposed/designed based on DL solutions to assist in the rapid diagnosis of COVID-19 from other types of respiratory infections~\cite{MeiX:2020, ArdakaniA:2020, BaiH:2020, LiL:2020, AfsharP:2020, WaheedA:2020, KhanA:2020, SinghD:2020, HuSh:2020}. There are, however, fewer works on developing DL-based models for segmentation and quantification of COVID-19 lesions. Segmentation models designed based on DL models are mainly developed based on CNNs. Such DL models are image-to-image networks containing an encoding path for extracting high-resolution features from input images and a decoding path for generating masks indicating the regions of interest.
The majority of the COVID-19 lesion segmentation models have been developed upon U-Net~\cite{RonnebergerO:2015} due to its superiority for the task of medical image segmentation~\cite{saeedizadeh2021covid, wang2020noise}. For example, using a U-Net architecture integrated with DenseNet blocks in the encoder path, Chaganti \textit{et al.}~\cite{ChagantiShO:2020} proposed a segmentation model to quantify lung abnormalities caused by COVID-19 from CT images. Similarly, Zhou \textit{et al.}~\cite{ZhouT:2020} proposed an enhanced U-Net segmentation model by incorporating spatial and channel attention mechanisms. Segmentation networks can be designed based on 2D or 3D CNNs to segment COVID-19 regions of infection either on slice-level or patient-level basis. It should be noted that 3D segmentation networks that can segment infections on the whole lung volumes are more desired from practical point of view. However, they need a large amount of 3D annotated lung volumes for efficient training and are more computationally expensive. Reference~\cite{wang2020noise} trained both 2D and 3D variants of the U-Net model for segmenting COVID-19 infections from chest CT scans of $558$ patients confirmed with COVID-19 pneumonia. According to their experiments, the 2D U-Net model achieved a Dice Similarity Coefficient (DSC) of $79.07$\% while 3D U-Net obtained DSC of $70.35$\%,  demonstrating that on small datasets, slice-level segmentation can achieve better results.

Generally speaking, to efficiently train a DL-based segmentation model, large amount of annotated CT images are required. However, due to nonuniform contrasts and irregular boundaries of COVID-19 lesions, providing pixel-level labeled datasets is challenging and expensive. Some research studies use segmentation methods that require less data for training to overcome the lack of sufficient pixel-level annotated CT images. Authors in Reference~\cite{fan2020inf} developed a semi-supervised segmentation framework that is trained only on $50$ annotated CT images of COVID-19 patients together with $1,600$ unlabeled CT images. Their proposed model segments the COVID-19 lung abnormalities under both one and two classes of infections. Integrating semi-supervised and few-shot learning methods, Reference~\cite{abdel2021fss} introduced a segmentation network that can learn from a few number of labeled chest CT scans. Authors proposed a dual-path network architecture for few-shot learning and incorporated an adaptive knowledge exchange module between the networks' paths to enhance the model's performance in segmenting COVID-19 lesions. Alternatively, one may choose to train a DL network with weak supervision through lower-quality labels, which are inexpensive and need less time to be generated. For instance, Laradji \textit{et al.}~\cite{laradji2021weakly} proposed a weakly supervised learning method by annotating a single pixel for any COVID-19 infection region on a CT image and achieved a DSC range of $68$\% - $75$\% on three public datasets. Their proposed labeling scheme reduced the annotation time for each infection region from $10$ - $15$ seconds to $1$ - $3$ seconds. Using some simple operations, Yao \textit{et al.}~\cite{yao2021label} synthesized COVID-19 regions of infection on chest CT scans of healthy people and generated fake image-label pairs for training their segmentation network with no labeled data. It is worth mentioning that although these studies suggest effective solutions in compensating the lack of sufficient labeled data, they cannot yield the same success of fully supervised learning methods with accurate labels. This study aimed to address this gap.

\vspace{.025in}
\noindent
\textbf{The $\CR$ Framework:}
Capitalizing on the above discussion, we proposed a deep convolutional neural network model, the so called $\CR$ framework, for segmenting COVID-19 legions from chest CT scans. A high-quality COVID-19 segmentation dataset containing $433$ CT slices from $82$ patients is also introduced.

Several comprehensive experiments were conducted to evaluate efficacy and limitations of the proposed $\CR$ model on 2D CT images and whole CT volumes, using both internal and external datasets. The results indicate that $\CR$ can efficiently segment COVID-19 regions of infection  from CT scans on both slice-level and patient-level basis. To cope with data hungry nature of deep AI models, a novel data augmentation method that generates synthetic CT images and infection masks by  inserting regions of infection  from COVID-19 infected CT scans to healthy CT images. In particular, based on the results of the comparison study (Table~\ref{tab:architec}), it can be observed that the $\CR$ network outperforms the standard U-Net networks considering all evaluation metrics. Furthermore, $\CR$ with less trainable parameters and computational complexity yields better results than the residual U-Net. Although adding the CPB module imposes some computational burden to the model and increases the number of trainable parameters from $6.32$ million to $8.75$ million, it can improve DSC, SEN, and MAE metrics while providing a similar SPC.
The results of the generalization experiment (Figures~\ref{fig:gnz_plt} and~\ref{fig:infrate}) illustrate that despite being trained only on CT images with infection, our proposed $\CR$ can optimally segment COVID-19 lesions on whole CT volumes. Indeed, by quantifying the regions of infection on 2D CT slices and summing them up for the entire lung volume, assuming that the changes over a thick CT slice are negligible, the model can approximate the patient-level infection rate in COVID-19 patients.
Finally, from the results of the experiment conduced to evaluate efficacy of the proposed synthetic data augmentation mechanism (Figure~\ref{fig:synth_im} and Table~\ref{tab:res-aug}), it can be observed that augmenting the training set using the proposed data augmentation method can improve the average DSC from $0.7599$ to $0.781$, the average SEN from $0.8104$ to $0.8308$, the average SPEC from 0.9953 to 0.9957, and the average MAE from $0.0074$ to $0.0067$. The experimental results indicate that concatenating the synthetic images generated by our data augmentation method will enrich the training set by introducing more variability to the training set, resulting in enhanced model performance.

Segmentation models are used as the first step of severity assessment and prognosis prediction of COVID-19 patients, which would help optimize resource allocation and patient management. Future directions include extending the $\CR$ for quantifying specific COVID-19 severity measures and integrating it into a hybrid deep learning model to detect high-risk COVID-19 patients and predict adverse outcomes based on CT images and clinical/laboratory information.


%
\section*{Acknowledgements}

This work was partially supported by the Natural Sciences and Engineering Research Council (NSERC) of Canada through the NSERC Discovery Grant RGPIN 2019 06966.

\section*{Data/Code Availability}

The datasets/codes generated and/or analyzed during the current study will be released publicly upon potential publication of the study.

\section*{Author Contributions Statement}

N.E. implemented the model with assistance of P.A. and SH.H. N.E. drafted the manuscript jointly with F.N., A.O. and M.J.R. supervised the clinical parts, F.N., K.N.P, and A.M. directed and supervised the study. All authors reviewed the manuscript.

\section*{Additional Information}
Competing Interests: Authors declare no competing interests.

\end{document}